\documentclass[10pt,conference]{IEEEtran}
\IEEEoverridecommandlockouts

\usepackage[T1]{fontenc}
\usepackage[utf8]{inputenc}
\usepackage{microtype}
\usepackage{amsmath,amssymb}
\usepackage{graphicx}
\graphicspath{{gfx/}{figures/}}
\usepackage{booktabs}
\usepackage{balance}
\usepackage{cite}
\usepackage{xcolor}
\definecolor{paperblue}{HTML}{0057FF}
\definecolor{citegreen}{HTML}{39FF14}
\usepackage[hypertexnames=false]{hyperref}
\usepackage{orcidlink}
\hypersetup{%
  colorlinks=true,
  linkcolor=paperblue,
  citecolor=citegreen,
  urlcolor=paperblue,
  pdfborder={0 0 0},
  pdftitle={From Quantum Shots to Training Data: Reorganizing Measurement Records in Quantum Machine Learning},
  pdfsubject={Measurement record aggregation for finite shot quantum machine learning},
  pdfkeywords={quantum reservoir computing, finite shot learning, measurement aggregation, time series forecasting}
}
\newcommand{\QAIBalancedTableRows}{%
Mackey--Glass & 0.098 & 0.190 & 0.484 & 0.439 & 0.586 & \textbf{0.430} \\
Santa Fe laser & 0.762 & 0.493 & 0.764 & 0.680 & 0.697 & \textbf{0.677} \\
R\"ossler & 0.096 & 0.054 & 0.414 & 0.379 & 0.580 & \textbf{0.367} \\
Sunspots & 0.387 & 0.474 & 0.553 & 0.517 & 0.558 & \textbf{0.515} \\
}

\begin{document}

\title{From Quantum Shots to Training Data:\\
Reorganizing Measurement Records in Quantum Machine Learning}

\hypersetup{pdfauthor={Markus Baumann; Maximilian Zorn; Thomas Gabor; Claudia Linnhoff-Popien; Jonas Stein}}
  \author{%
  \IEEEauthorblockN{%
  Markus Baumann\IEEEauthorrefmark{1}\orcidlink{0009-0007-3575-1006}\thanks{Corresponding author: \href{mailto:markus.baumann@campus.lmu.de}{markus.baumann@campus.lmu.de}.},
  Maximilian Zorn\IEEEauthorrefmark{1}\orcidlink{0009-0006-2750-7495},
  Thomas Gabor\IEEEauthorrefmark{2}\orcidlink{0000-0003-2048-8667},
  Claudia Linnhoff-Popien\IEEEauthorrefmark{1}\orcidlink{0000-0001-6284-9286},
  and Jonas Stein\IEEEauthorrefmark{1}\orcidlink{0000-0001-5727-9151}}
  \IEEEauthorblockA{\IEEEauthorrefmark{1}\textit{QAR-Lab, Department of Computer Science, LMU Munich, Munich, Germany}\\
  \IEEEauthorrefmark{2}\textit{Department of Computer Science, University of Exeter, Exeter, United Kingdom}}
  }

\maketitle
\bstctlcite{BSTcontrol}

\begin{abstract}
Between keeping every shot as a noisy training example and averaging all shots
into one clean feature lies an entire spectrum of data organizations that
quantum machine learning usually leaves implicit. Shot grouping makes this
choice explicit by partitioning a fixed measurement record into disjoint groups
and averaging within each group, tuning smoothly between the two conventions
with a single validated parameter and no additional quantum executions. We evaluate
the method on chaotic synthetic benchmarks and on real-world data under strict
execution budgets. At the balanced allocation, intermediate grouping lowers
mean error across the evaluated tasks, with supported improvements on every
task. Matched controls show that the gain acts as
correctly scaled regularization of the classical readout. Unmitigated runs on
two superconducting processors show positive mean differences on every task.
Shot grouping therefore offers a practical way to improve finite-shot quantum
learning at no added quantum cost.
\end{abstract}


\section{Introduction}
\label{sec:introduction}

Every quantum learning experiment ends the same way: a circuit runs many
times, and each run returns one measured bitstring. Before anything can be
learned, this pile of shots must become training data. The step looks like
bookkeeping, and it is usually treated that way. Almost every pipeline averages
all shots of a time step into a single expectation-value estimate and moves on.
This study examines the statistical consequences of the usual practice of
averaging all shots at each time step.

This convention is well motivated because expectation values are the quantities
predicted by theory and estimated in experiments. A learning pipeline, however,
does not consume observables.
It consumes training examples, and it cares about two things only: how many
there are and how noisy each one is. The moment a measurement record becomes
training data, a physics convention quietly turns into a statistical decision:
spend all measurements on making one example clean, or make several examples
that are merely clean enough. No single choice can suit every task, budget,
and noise level.

Averaging everything is one end of a dial. It buys the cleanest possible
feature vector but pays with scarcity. Thousands of circuit executions collapse
into a handful of training examples, and too few examples relative to the
number of features destabilizes regression~\cite{hsu2012random}. The opposite
end, keeping every shot as its own example, buys abundance but pays with noise,
and noisy features systematically shrink fitted
coefficients~\cite{fuller1987measurement}. Between these extremes lies an
entire range of aggregations. Nothing in the record forces either convention,
and the record does not care which one tradition prefers.

Quantum reservoir computing (QRC) is a clean place to study this
dial~\cite{fujii2017harnessing,nakajima2019boosting,mujal2021opportunities}. A
fixed quantum circuit transforms an input sequence into measured features, and
only a classical readout is trained. The circuit, the targets, and the
measurement record can therefore be held completely fixed while the aggregation
alone changes. Whatever happens to the forecast is attributable to the
aggregation because nothing else moved.

\begin{figure}[!t]
  \centering
  \includegraphics[width=\columnwidth]{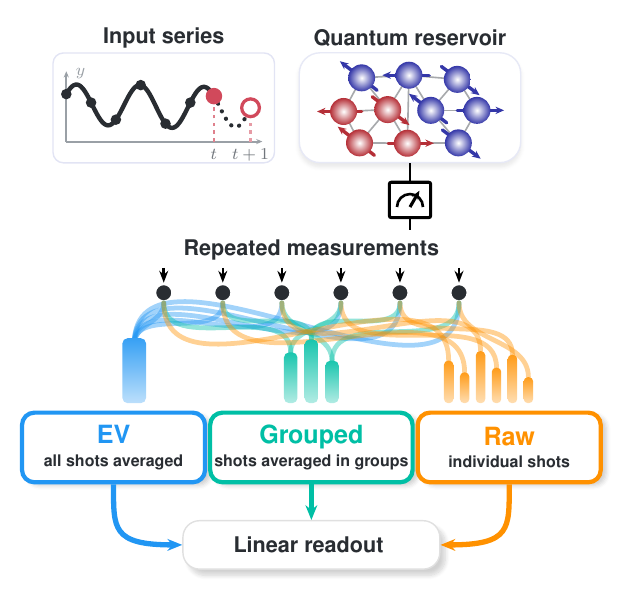}
  \caption{Three ways to turn one shot record into training data. A fixed quantum
  reservoir is measured many times per step, and only the classical step changes:
  EV averages all shots, Raw keeps each shot, and Grouped averages disjoint groups.
  All feed the same linear readout.}
  \label{fig:organization}
\end{figure}

We compare three representations of one record
(Fig.~\ref{fig:organization}). The expectation-value representation (EV)
averages all measurements at each time step. Raw uses each measurement
separately. Grouped averages disjoint subsets, and its aggregation scale $k$,
the number of shots per averaged training view, slides smoothly from $k=1$
(Raw) to $k=N$ (EV).

Under the matched protocol and fixed execution budgets, the aggregation scale
materially affects test error. At the balanced allocation, an intermediate
aggregation lowers mean test error relative to full averaging on all four tasks,
with paired confidence intervals above zero in every case. The best aggregation
scale depends on the record, so no single convention is appropriate. Because
the improvement arises entirely from classical post-processing, it requires no
additional quantum executions. Matched controls support a
covariance-regularization explanation for the classical readout, supplied
automatically by the record's own fluctuations. When individual shot outcomes
have been retained, the same post-processing can be applied retroactively to an
archived record.

Adaptive shot aggregation treats the number of shots per training view as a
validation-selected hyperparameter of the classical readout. We map
when the choice matters through three budget allocations of time steps and
shots. We explain why it works through matched controls that isolate the
regularization mechanism. We test where it transfers through unmitigated
experiments with a separate hardware reservoir on two superconducting
processors.

\section{Related Work}
\label{sec:related}

Finite sampling can be addressed at three stages of a quantum learning
pipeline: before execution, by allocating the measurement budget; after
averaging, by correcting expectation-value estimates; or at the measurement
record itself, by deciding how repeated outcomes become training examples. Our
method focuses on this third stage.

A brief instance of subset averaging appears in Cemin \emph{et al.}~
\cite{cemin2024channels}, where it is used as an implementation heuristic to
enlarge the training set for reduced quantum channel learning. We develop this
idea into a tunable procedure: the group size spans the full range from
individual shots to complete expectation-value averaging, is selected jointly
with readout regularization, and is evaluated under fixed execution budgets.
For linear readouts, grouping also has an exact regularization interpretation,
connecting it to classical work on training with noisy inputs~
\cite{bishop1995training}.

Complementary work addresses the other stages of finite-shot learning.
Resource-aware optimization allocates measurements before execution~
\cite{moussa2023resource}. After averaging, noise-aware training rules~
\cite{ahmed2025optimal} and variance regularization~
\cite{kreplin2024reduction} reduce the effect of sampling noise, while
fundamental limits for learning from finitely sampled physical systems have
also been studied~\cite{hu2023tackling}. Shot grouping instead reorganizes an
existing measurement record. It requires no additional quantum executions and
can be applied retrospectively whenever individual outcomes have been retained.

Within QRC, prior work has examined how noise, dissipation, measurement, finite
Hilbert-space dimension, and input transformations shape temporal features~
\cite{kubota2022noise,palacios2024role,kubota2023temporal,
martinezpena2023finite,mujal2023timeseries,suzuki2022natural,
sannia2024dissipation,govia2022nonlinear,martinezpena2023capacity,
kalfus2022hilbert,fry2023noiseinduced,govia2021singleoscillator}. We hold the
reservoir and measurement record fixed and change only how repeated outcomes
are presented to the classical readout. Shot grouping is therefore
complementary to these approaches and is not tied to a particular reservoir
design.

\section{Method}
\label{sec:method}

\subsection{Adaptive shot aggregation}

Write $x_t^{(n)}$ for the features produced by shot $n$ at time $t$. We split
the $N$ shots of each step into $G=N/k$ groups of equal size $k$ and average
within each group,
\begin{equation}
  x_t^{(g)}=\frac{1}{k}\sum_{n\in g}x_t^{(n)}, \qquad g=1,\dots,G.
  \label{eq:grouping}
\end{equation}
Every group is a partly averaged view of the same time step, and all views
share the same target $y_t$. At test time we average the $G$ predictions of a
step, which is identical to predicting once from the averaged features. The two
methods are therefore scored in exactly the same way, so any difference between
EV and Grouped is created during training, not at test time.

The shot index carries no physical meaning, because shot $n$ at one step and
shot $n$ at the next come from unrelated circuit runs. We therefore shuffle the
shots freshly at every time step before forming groups, and we repeat the whole
analysis over several random groupings.

\subsection{Fixed execution budget and training}

Evaluating $T$ time steps with $N$ shots per step costs $TN$ circuit
executions. We hold this budget fixed at ten thousand executions and study
three ways of spending it, $(T,N) = (250,40)$, $(500,20)$, and $(1000,10)$,
ranging from fewer time steps with more shots to more time steps with fewer
shots. For each setting we try every group size $k$ that divides $N$, which
includes Raw at $k=1$ and EV at $k=N$.

Selection and testing respect time order. After a short washout, the earlier
part of each series is the selection data used to choose settings, and the
final fifth is held out as the test block and left untouched until the end. On
the selection data we pick the group size and ridge strength $(k,\lambda)$ by
forward-chaining validation. In three successive rounds we fit on an earlier
stretch and score on the stretch that immediately follows, so training always
precedes validation in time, and each candidate pair is ranked by its
validation error averaged over the three rounds. The best pair is then frozen,
the readout is refitted on the whole selection data, and the test block is
scored once. Normalization and feature scaling are computed from training data
only, so the test block never influences any choice.

\begin{figure}[!t]
  \centering
  \includegraphics[width=\columnwidth]{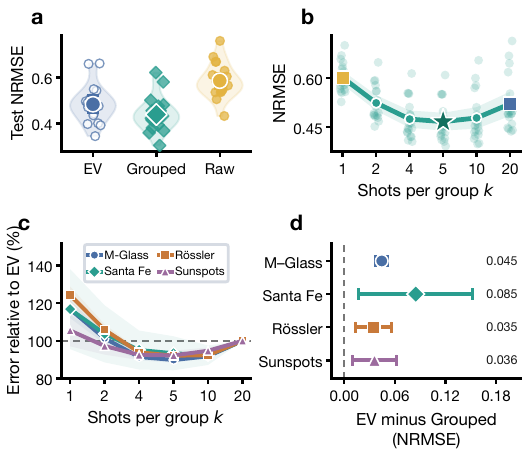}
  \caption{Adaptive aggregation under the shared protocol. (a)~Distributions of
  EV (blue), Grouped (teal), and Raw (yellow) errors across twenty
  Mackey--Glass trajectories, with large markers and bars showing means and
  95\% intervals. (b)~The error is
  U-shaped in the aggregation scale $k$, lowest at an intermediate value (star is
  the validated choice). (c)~The same dip below EV appears in the task means.
  (d)~Paired EV-minus-Grouped improvements on the test blocks, with markers and
  bars showing means and 95\% intervals across independent trajectories or
  windows. Positive values favor Grouped. Panels (b,c) use development data,
  (a,d) test blocks.}
  \label{fig:operating}
\end{figure}

\subsection{Why grouping regularizes the readout}
\label{sec:mechanism}

Full averaging gives the readout one relatively stable feature vector per time
step. Raw gives it many noisier vectors. Grouping lies between them, providing
several partly averaged views of the same time step with the same target.

Stable feature combinations should predict similarly across these views,
whereas shot-sensitive combinations do not. Fitting every view to the same
target therefore discourages reliance on finite-shot fluctuations.

For a linear readout with squared loss, this effect is exact. Let $\bar{x}_t$ be
the average of the grouped views, and let
$C_{k,t}=\tfrac{1}{G}\sum_g
(x_t^{(g)}-\bar{x}_t)(x_t^{(g)}-\bar{x}_t)^\top$
be their covariance around that average. For any intercept $b$ and weights $w$,
\begin{equation}
  \frac{1}{G}\sum_{g=1}^{G}\bigl(y_t-b-w^\top x_t^{(g)}\bigr)^2
  =\bigl(y_t-b-w^\top\bar{x}_t\bigr)^2+w^\top C_{k,t}w.
  \label{eq:loss-decomposition}
\end{equation}
The first term is the EV loss. The second penalizes feature combinations that
fluctuate across views. Grouped training is therefore EV training with a
regularizer obtained from observed shot fluctuations. The identity requires
neither Gaussian noise nor independent time steps.

On a fixed common feature scale, let $S_{\mathrm{shot},t}$ be the $1/N$
covariance of the $N$ shot-level features. Before representation-specific
standardization, averaging over random partitions $\mathcal{P}$ gives
\begin{equation}
  \mathbb{E}_{\mathcal{P}}[C_{k,t}]
  =\frac{N-k}{k(N-1)}S_{\mathrm{shot},t}.
  \label{eq:penalty-scale}
\end{equation}
The multiplier is one at $k=1$ and zero at $k=N$. Thus, $k$ mainly controls
regularization strength: smaller groups retain more shot variation, while
larger groups average it away. A finite partition can also change the realized
penalty, especially when only a few groups are available.

This motivates covariance-regularized expectation-value training (Cov-EV),
which retains EV features and adds a penalty estimated from shot fluctuations.
Its covariance and ridge strengths are selected on the same validation data,
independently of Grouped's $k$. Similar performance from Grouped and Cov-EV
would support covariance regularization as the mechanism behind the gain.

As a check on our implementation, at a fixed group size and ridge strength we
compute the Grouped fit in two mathematically equal ways. The first fits ridge
directly on all grouped training views. The second uses the two-term form of
Eq.~\eqref{eq:loss-decomposition}, fitting on the mean view with the added
covariance penalty. With the same feature standardization, the two agree to
numerical precision, which confirms that the decomposition is implemented
correctly.

\section{Experimental Setup}
\label{sec:setup}

We run two studies. The first is a controlled simulation on four forecasting
tasks. The second repeats the core comparison on real quantum hardware. In both,
every aggregation choice reads the very same measurement record, so any
difference in accuracy comes only from how the shots are organized and not from
running the circuit differently.

\subsection{Simulation reservoir and evaluation protocol}

The simulated reservoir is a fixed six-qubit circuit with two shallow data
reuploading layers~\cite{perezsalinas2020data}. Each layer applies
input-dependent $R_y$ and $R_z$ rotations with fixed per-qubit offsets,
nearest-neighbor $R_{zz}$ couplings, and fixed non-Clifford mixing rotations.
For each circuit setting, all angles are sampled once and then held fixed;
three frozen settings are used to assess sensitivity to this draw. Each
computational-basis shot yields 11 features: six local observables $Z_i$ and
five nearest-neighbor products $Z_iZ_{i+1}$. A leaky integrator with leak rate
$0.2$ and a ten-step lag window provide temporal memory, and a ridge-regression
readout predicts the next value.

We evaluate one-step-ahead forecasting on four time series: the synthetic
Mackey--Glass~\cite{mackey1977oscillation} and
R\"ossler~\cite{rossler1976equation} systems, and the measured Santa Fe laser
intensity~\cite{gershenfeld1993future} and monthly sunspot
number~\cite{sidc2026sunspots}. All tasks use the same reservoir architecture,
feature construction, temporal processing, scaling, validation procedure, and
execution budgets. The same selection rule is applied separately to each task;
no task-specific changes are made to the reservoir or preprocessing pipeline.

EV, Grouped, Raw, and Cov-EV are derived from the same finite-shot measurement
record. EV averages all shots at each time step, Raw retains individual shots,
Grouped averages shots within disjoint groups, and Cov-EV uses the fully
averaged EV features with a penalty estimated from shot-level covariance. Two
additional controls use different feature sources: an ideal reservoir without
finite-shot sampling noise and a classical baseline that applies the same
temporal processing and readout directly to the input series. These controls
distinguish the effect of aggregation from the effects of finite sampling and
of the reservoir transformation itself.

The trajectory or time window is the unit of uncertainty estimation.
Mackey--Glass uses twenty independently generated trajectories, each with a
held-out test block. R\"ossler uses five trajectories, and each real-world task
uses five windows with disjoint test blocks.
Within each unit,
we repeat the analysis over three circuit settings, three measurement draws,
and three random groupings where applicable. These repeats are averaged within
the unit before paired confidence intervals are formed across the independent
units for that task.
Grouped views, neighboring time steps, and repeated evaluations of the same
sequence are not treated as independent observations, because they share
measurements or are coupled by the leaky integrator and overlapping lags.

Model selection uses no test data. After a short washout, the final 20\% of each
trajectory or window is reserved as an untouched test block. The preceding
data are used in three expanding-window validation splits, each trained on the
available past and evaluated on the immediately following block. The group
size $k$ and ridge strength $\lambda$ are selected from these validation
results. The readout is then refitted on all data preceding the test block and
evaluated once on the reserved block. For representations with multiple views
at one time step, the corresponding predictions are averaged to produce one
forecast, so all methods are scored on the same targets.

The primary budget uses $T=500$ time steps and $N=20$ shots per step, giving
10,000 circuit executions. We also test $(T,N)=(250,40)$ and $(1000,10)$,
which spend the same budget with different balances between observed time
steps and shots per step. All reservoir parameters, data-generation settings,
window origins, and selection rules are fixed and stored before any test block
is evaluated.

\subsection{Hardware setup}
The hardware study deliberately uses a different reservoir. It is a fixed
ten-qubit circuit with one input qubit, nine memory qubits, a single ring of
entangling gates, and ten local outputs, mapped to selected qubits on the
\emph{ibm\_fez} and \emph{ibm\_marrakesh} processors and held fixed across all
four tasks. Its purpose is to check whether the same aggregation procedure still
helps on a second circuit and a different set of measured features. It is not a
hardware implementation of the six-qubit simulator and not a comparison of
qubit counts. As in
simulation, every method receives the same circuit outputs and only the
classical grouping of shots changes. For each task and processor we take one
acquisition of $400$ shots at each of $500$ time steps and split it into twenty
ordered, disjoint records of $20$ shots. The resulting intervals describe how
much the result varies among the measurement blocks of that single acquisition.
They do not stand in for independent devices or separate calibration periods.

\begin{figure}[!t]
  \centering
  \includegraphics[width=\columnwidth]{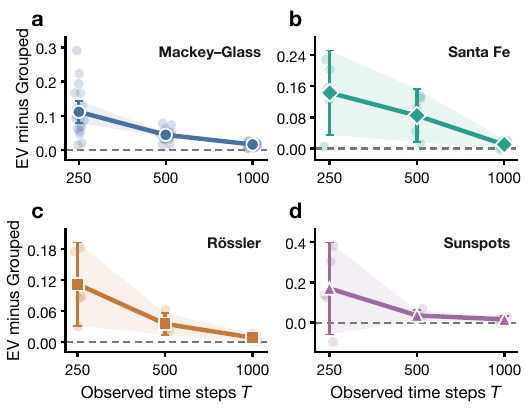}
  \caption{When aggregation helps most, at a fixed budget. Each panel shows the
  improvement of Grouped over EV, measured as $\mathrm{NRMSE}_{\mathrm{EV}} -
  \mathrm{NRMSE}_{\mathrm{Grouped}}$, so values above zero favor Grouped. Moving
  right spends the budget on more steps and fewer shots. The gain is largest with
  many shots and few steps. Markers and bars are means with $95\%$ intervals.}
  \label{fig:regimes}
\end{figure}

\section{Evaluation}
\label{sec:evaluation}

We report accuracy as test NRMSE, a normalized error for which lower is better
and where one means predicting the mean. Three ways of turning the shot record
into features are compared throughout. EV averages all shots of a step into one
clean feature vector. Raw keeps every shot as its own noisy example. Grouped is
our method and sits between them. A fourth entry, Cov-EV, is a control that adds
an independently tuned covariance penalty to plain averaging. We summarize the
benefit of Grouped as the paired difference EV minus Grouped, so a positive
value means Grouped is more accurate. Each difference comes with a $95\%$
confidence interval. An interval above zero provides evidence of a positive
mean improvement; an interval containing zero is inconclusive. The balanced
setting spends the fixed budget on $500$ time steps with $20$ shots each.

\subsection{Balanced setting}

At the balanced budget, Grouped lowers mean NRMSE relative to EV on all four
tasks (Table~\ref{tab:balanced}, Fig.~\ref{fig:operating}). For Mackey--Glass,
the primary estimate uses twenty independent held-out trajectories. The improvement is
$0.0445$ on Mackey--Glass ($95\%$ CI $[0.0372, 0.0519]$), $0.0848$ on Santa Fe
laser ($[0.0174, 0.1522]$), $0.0349$ on R\"ossler ($[0.0139, 0.0559]$), and
$0.0360$ on sunspots ($[0.0103, 0.0616]$). Every interval stays above zero.

\begin{table}[t]
\caption{Mean test NRMSE at the balanced setting. Repeated measurement trials
are averaged within independent trajectories or windows. Lower
values are better.}
\label{tab:balanced}
\centering
\scriptsize
\setlength{\tabcolsep}{1.7pt}
\begin{tabular}{lrrrrrr}
\toprule
Task & Class. & Ideal & EV & Grouped & Raw & Cov-EV \\
\midrule
\QAIBalancedTableRows
\bottomrule
\end{tabular}
\end{table}

Santa Fe laser shows the largest effect. Grouped reduces the error from $0.764$
to $0.680$, and all five disjoint test windows favor Grouped. When $k$ and the
ridge strength are tuned together, the search lands on $k=10$ most often and
on an intermediate scale, neither Raw nor EV, in $94\%$ of repeated trials.
Cov-EV reaches $0.677$, almost the same as Grouped, which already suggests that
the gain is a regularization effect rather than new physical information. The
gain also depends on how the budget is spent. On Santa Fe it shrinks from
$0.143$ with $40$ shots per step to $0.010$ with $10$ shots per step because
grouping has more shots to reorganize when shots are plentiful.

We also compare an ideal noise-free reservoir with a classical baseline that
uses identical preprocessing. The quantum reservoir is worse on Mackey--Glass
and sunspots, better on Santa Fe, and inconclusive on R\"ossler. Our claims are
therefore about organizing a chosen reservoir record, not quantum advantage.
The preferred aggregation scale varies across tasks, trajectories, circuit
settings, measurement draws, and grouping assignments, and EV is occasionally
best. On a fixed feature scale, Eq.~\eqref{eq:penalty-scale}
clarifies the role of $k$: it mainly sets the strength of the covariance
penalty, while a finite random grouping can also change the realized penalty.
Raw is less accurate than
Grouped on all four task means.

\subsection{Trading time steps against shots}

Figure~\ref{fig:regimes} compares three ways of spending the fixed budget. With
$40$ shots at each of $250$ time steps, Grouped improves every task by
$0.112$ to $0.170$. The interval stays above zero for Mackey--Glass, Santa Fe
laser, and R\"ossler, while the sunspot result varies across windows. When the
same budget is spread over $1000$ time steps with only $10$ shots each, the
gains shrink to between $0.008$ and $0.016$, and the interval stays above zero
only for Mackey--Glass and sunspots.

\subsection{Why the gain is regularization}

\begin{figure}[t]
  \centering
  \includegraphics[width=\columnwidth]{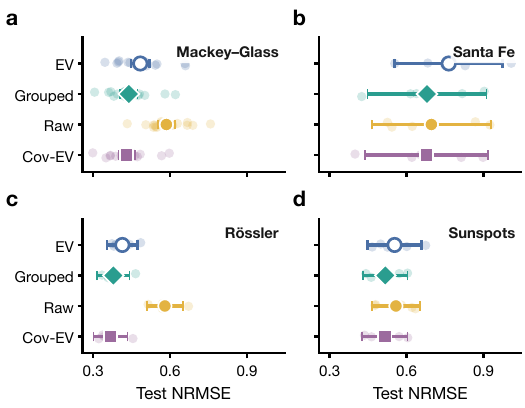}
  \caption{Controls at the balanced setting, lower is better. On every task
  Grouped (teal diamond) and Cov-EV (purple square) are lowest and nearly equal,
  while EV (open circle) and Raw (yellow) trail. Their near-match supports a
  readout-regularization explanation. Faint points are the independent unit means,
  markers and bars are means with $95\%$ intervals.}
  \label{fig:controls}
\end{figure}

Cov-EV keeps EV features and adds a penalty based on how the shots fluctuate
together. At the balanced setting its mean error is $0.0021$ to $0.0116$ NRMSE
below Grouped. The Mackey--Glass and R\"ossler intervals exclude zero. This near-match
supports the regularization mechanism in Eq.~\eqref{eq:loss-decomposition}
rather than extra device information (Fig.~\ref{fig:controls}). Raw is less
accurate than EV on Mackey--Glass and R\"ossler and inconclusive on the two
real-world datasets. At fixed $k$, the exact two-term loss reproduces the
Grouped ridge fit to numerical precision. On a fixed feature scale,
Eq.~\eqref{eq:penalty-scale} shows how the induced penalty changes
with $k$, whereas Cov-EV tunes the full shot-covariance penalty directly.

\subsection{Transfer to hardware}

\begin{figure}[!b]
  \centering
  \includegraphics[width=\columnwidth]{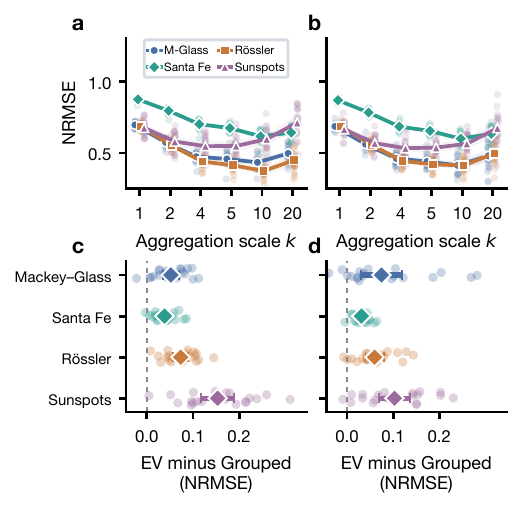}
  \caption{Transfer to two unmitigated processors. (a,b)~On \emph{ibm\_fez} and
  \emph{ibm\_marrakesh} the error is again U-shaped in the aggregation scale $k$.
  (c,d)~The improvement of Grouped over EV is positive on all four tasks and both
  devices. One $T=500$ acquisition of $400$ shots per step is split into twenty
  disjoint $N=20$ records, and intervals describe variation within that single
  acquisition.}
  \label{fig:hardware}
\end{figure}

Figure~\ref{fig:hardware} reports the unmitigated hardware results. Grouped
improves every task, with mean gains from $0.038$ to $0.153$ NRMSE on
\emph{ibm\_fez} and from $0.031$ to $0.103$ on \emph{ibm\_marrakesh}. On Santa
Fe laser, the error falls from $0.750$ to $0.711$ on \emph{ibm\_fez} and from
$0.744$ to $0.713$ on \emph{ibm\_marrakesh}. Grouped is more accurate in $19$
of $20$ measurement blocks on each processor.

\section{Discussion}
\label{sec:discussion}

A quantum experiment produces a measurement record, not a ready made training
set. Combining repeated shots into feature vectors is therefore part of the
learning algorithm. EV averages all $N$ shots at each step. Adaptive shot
aggregation instead averages groups of size $k$, from Raw at $k=1$ to EV at
$k=N$, without changing the circuit or record. The settings are fixed before
testing, and gains arise during training at no additional quantum cost.

The choice matters most when many shots are collected at relatively few time
steps. Full averaging then spends a large measurement budget on a small number
of highly precise examples. Grouping creates moderately denoised views from
the same shots. Their differences reveal which feature directions are stable
under finite shot fluctuations. With fewer shots per step, the freedom to
reorganize the record and the benefit both shrink. The aggregation scale
should therefore be validated rather than fixed by convention.

For a linear readout, this effect has an exact interpretation.
Equation~\eqref{eq:loss-decomposition} separates the grouped loss into the EV
loss and a quadratic covariance penalty. Since every view shares the same
target, the penalty discourages large weights on feature combinations that
vary between groups. On a common feature scale, Eq.~\eqref{eq:penalty-scale}
shows that smaller groups produce stronger regularization and $k=N$ recovers
EV.

The Cov EV control supports this explanation. It keeps the fully averaged
features but adds a tuned covariance penalty estimated from shot fluctuations.
It stays close to Grouped on every task and reaches slightly lower mean error
overall, pointing to a shared mechanism. Cov EV tunes the penalty directly,
while Grouped uses strengths produced by the available group sizes. Grouping
is therefore a simple route to this regularization, but not the only one.

Individual shot outcomes should be retained. Once a record has been reduced to expectation
values, alternative aggregation scales cannot be reconstructed. Keeping the
full record preserves this option without repeating the quantum experiment.

The effect transfers to a separate ten qubit circuit run without mitigation on
two superconducting processors, where Grouped improves on EV across all tasks.
The hardware intervals measure blocks within one acquisition per task and
processor, not reproducibility across runs, calibration states, or days.
Independent hardware campaigns, other reservoirs, and richer readouts are the
next tests. More broadly, finite shot processing is part of the learning
algorithm.

\section{Conclusion}
\label{sec:conclusion}

Turning a shot record into training data has always involved a choice, and the
field has usually answered it by averaging everything. Choosing the aggregation
scale by validation lowers mean test error on all four forecasting tasks at the
balanced budget, with paired confidence intervals above zero in every task.
The hardware study shows positive
mean differences on both superconducting processors, and the method requires no
additional quantum cost. Because every measurement-based reservoir produces such
a record, the choice is available to all of them. The underlying question, how
to turn finitely sampled noisy measurements into training data, reaches beyond
reservoirs to finite-shot learning as a whole.

\appendices
\section{Reproducibility}
The supplement and repository provide the complete circuit definitions and
parameters, physical qubit mappings, transpilation settings, backend and
calibration metadata, hyperparameter grids, normalization order,
confidence-interval construction, random seeds, data-window origins, raw and
derived records, and the code used to reproduce every figure and table. The
\ifdefined\ANON
anonymous artifact is available at
\href{https://anonymous.4open.science/r/split-ensemble-qrc-E688/README.md}{anonymous artifact repository}.
\else
named artifact is available at
\href{https://github.com/eybmits/adaptive-shot-aggregation}{github.com/eybmits/adaptive-shot-aggregation}.
\fi

\vspace*{-10pt}
\balance
\bibliographystyle{IEEEtranDoi}
\bibliography{bstcontrol,references}

\end{document}